\documentstyle[prl,aps,epsfig]{revtex}

\newcommand{\beq}{\begin{equation}}
\newcommand{\eeq}{\end{equation}}
\newcommand{\beqas}{\begin{eqnarray*}}
\newcommand{\eeqas}{\end{eqnarray*}}
\renewcommand{\vec}[1]{\mbox{\boldmath$#1$}}

\begin{document}
\twocolumn[\hsize\textwidth\columnwidth\hsize\csname 
@twocolumnfalse\endcsname

\title{Rotationally driven gas of inelastic rough spheres}

\author{Raffaele Cafiero (1), Stefan Luding (2), 
        and Hans J\"urgen Herrmann (1,2)}
\address{
(1) P.M.M.H., \'Ecole Sup\'erieure de Physique et de Chimie 
 Industrielles (ESPCI), \\ 
 10, rue Vauquelin, 75251 Paris CEDEX 05, FRANCE \\
(2) Institute for Computer Applications 1, 
 Pfaffenwaldring 27, 70569 Stuttgart, GERMANY\\}

\maketitle

\begin{abstract}
We study a two-dimensional gas of inelastic rough
 spheres, driven on the rotational degrees of freedom.
Numerical simulations are compared to mean-field (MF) predictions 
with surprisingly good agreement for strong coupling of
rotational and translational degrees of freedom  
-- even for very strong dissipation in the translational degrees.
Although the system is spatially homogeneous, 
the rotational velocity distribution is essentially Maxwellian. 
Surprisingly, the distribution of tangential velocities is
strongly deviating from a Maxwellian.
An interpretation of these results is proposed, as well as 
a setup for an experiment.
~
\end{abstract}

\narrowtext
\vskip2pc]

Systems of hard spheres have a long-standing history as a basic model for 
gases, liquids, and e.g.~glasses.  When dissipation is added, one has the
minimal model for granular materials and it is only a small step to
include also the rotational degrees of freedom via a tangential 
interaction at contact.
Granular materials belong to the fascinating world of non-linear, 
dissipative, non-equilibrium systems 
\cite{jaeger96b,herrmann98,poschel00,mecke00}, 
whose interest is due to their practical importance and due to 
the theoretical challenges they represent.
Granular media are collections of macroscopic 
particles with rough surfaces and dissipative, frictional interactions.
Numerical simulations are an established tool to complement advanced
theoretical approaches and difficult experimental studies.

In order to study systems of rough spheres, kinetic theories
have been extended to (weak) dissipation and friction
\cite{jenkins85b,lun87}.  Alternative, more recent approaches are 
based on a pseudo-Liouville operator formalism 
\cite{huthmann98,luding98d,herbst00,cafiero00b}
and are less general in the sense that they assume homogeneity and 
Maxwellian velocity distributions in order to arrive at a mean field 
(MF) description of systems with rotational degrees of freedom.
Either the system is left undisturbed \cite{huthmann98,luding98d,herbst00}
and thus cools continuously,
or a ``driving'' can be applied, i.e.~energy is fed into the system
in order to reach a steady-state situation.

The typical driving of a granular material, in both experiment and 
simulation, can be realized by moving walls 
\cite{herrmann98} which lead to rather localized input of energy. 
Alternatively, the system can 
be driven by a global homogeneous, random energy source 
in different variations \cite{cafiero00,puglisi99,bizon00,noije98,noije98b}. 
Depending on the experimental setup, energy can be given either to
translational degrees of freedom or to the rotational ones, or to
both. The first case catched most of the attention -- reason 
enough to change the focus and feed rotational energy instead
of translational.
In the experiment, translational energy input
was applied for special boundary conditions and a variety of 
interesting experimental results were obtained just recently
\cite{kudrolli97,olafsen98,olafsen99,losert99,losert99b}.
One can obtain a gas and a liquid state, together with 
dense, solid-like clusters which form due to dissipation.

The dynamics of the system is usually assumed to be
dominated by two-particle collisions which are modeled by their
asymptotic states: A collision is characterized by the velocities
before and after the contact, and the contact is assumed to be 
instantaneous.
In the simplest model, one describes inelastic collisions by a
normal restitution coefficient $r$ only, i.e.~the negative ratio
between the normal velocities after and before the collision. 
However, since surface roughness and friction are important  
\cite{huthmann98,luding98d,herbst00,cafiero00b,luding95b,mcnamara98}, 
one should allow for an exchange of translational and rotational energy.
In the simplest approach 
\cite{jenkins85b,huthmann98,mcnamara98}, 
surface roughness is accounted for by a constant tangential 
restitution coefficient $r_t$, which is defined in analogy 
to $r$ in the tangential direction. A more realistic friction law 
involves the Coulomb friction coefficient 
\cite{herbst00,cafiero00b,walton86,foerster94,luding98c}, so that
the tangential restitution will depend on the collision angle.
Constant tangential restitution is recovered 
in the limit of perfect friction.

In this Letter, we will focus on a system of such perfectly rough 
particles, where only the rotational degrees of freedom are coupled 
to a homogeneous driving. 
Such a situation could
correspond, for example, to a gas of rough magnetic particles 
subject to a rapidly varying, homogeneous, magnetic field. 
Besides a possible experimental application, we believe that
this study is interesting in itself, since a correct modeling 
of the driving mechanism is of great importance for a theory
of granular gases to describe realistic experimental situations.

\begin{figure}[htb]
\begin{center}
\epsfig{file=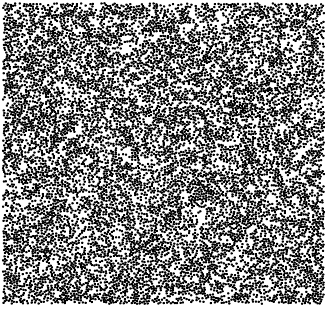,height=4.0cm,angle=0}
\end{center}
\caption{Snapshot of the particle distribution in the
steady state for a system of $N=11025$ particles,
$\nu=0.34$, $r_t=1$, and $r_n=0.1$.}
\label{fig1}
\end{figure}

The model consists of $N$ three-dimensional spheres with
radius $a$ and mass $M$,
interacting via a hard-core potential and confined to a 2D
plane of linear extension $L$, with periodic boundary conditions. 
The degrees of freedom are the positions $\vec r_{i}(t)$
the translational velocities $\vec v_{i}(t)$ and the rotational
velocities $\vec \omega_{i}(t)$ for each sphere numbered by 
$i=1, \ldots, N$. 
When two particles $1$ and $2$ 
collide, their velocities after collision are related to the 
velocities before collision, through a collision 
matrix which is derived from the linear and angular momentum 
conservation laws, energy/dissipation balance.  The magnitude
of dissipation is proportional to the quantities $1-r^2$
and $1-r_t^2$, while the strength of the coupling between
rotational and translational motion is connected to $1+r_t$,
where the normal restitution $r$ varies between $1$ (elastic) 
and $0$ (inelastic) and the tangential restitution $r_t$ 
varies between $-1$ (smooth) and $+1$ (rough), corresponding to
zero and maximum coupling, respectively  
\cite{jenkins85b,huthmann98,mcnamara98}. 
For a typical steady-state configuration (energy input is specified
below), see Fig.\ \ref{fig1}.

In order to feed energy,
the system is agitated each time interval $\Delta t=f_{\rm dr}^{-1}$, 
with a driving rate $f_{\rm dr}$.  Here, we will not 
apply driving frequencies much higher than the collision 
rate $\Omega$, but will use driving frequencies around 
$100$\,s$^{-1}$, comparable to $\Omega$.  This is rectified, since
numerical checks with strongly different values of $f_{\rm dr}$ 
lead to a similar behavior of the system even for driving frequencies 
lower than, but of the same order as $\Omega$, provided that a stationary 
state is reached. 
The translational velocity remains unchanged, but the
angular velocity $\omega_i$ of particle $i$ is modified at each time 
of agitation $t$ so that
\begin{eqnarray}
\omega' _i(t) & = & \omega_i(t)+r_i \, \omega_0 ~, \nonumber 
\label{eq:vxy}
\end{eqnarray}
where the prime on the left hand side indicates the value after 
the driving event. Due to the two-dimensionality of the system,
we apply the driving force only to the $z$-direction, so that
the scalar $\omega$ is to be understood as the $z$-component
of $\vec \omega$. 
$\omega_0$ is a reference angular velocity (in this study we use 
$\omega_0=2.4\,\,10^{-4}$\,s$^{-1}$) which allows, 
with $v_0=a \omega_0=2.4\,\,10^{-7}$\,m\,s$^{-1}$, where
 $a=10^{-3}$\,m, to
define the dimensionless translational and rotational particle temperatures
$T_{\rm tr}=E_{\rm tr}/(N T_0)$ and $T_{\rm rot}= 2 E_{\rm rot}/(N T_0)$,
with the translational energy $E_{\rm tr} = (M/2) \sum_{i=1}^N \vec{v}_i^2$,
the rotational energy $E_{\rm rot} = (qMa^2/2) \sum_{i=1}^N \vec{\omega}_i^2$,
and the reference temperature $T_0=M v_0^2$.
The variance of the uncorrelated Gaussian random numbers 
$r_i$ (with zero mean) can now be interpreted 
as a dimensionless driving temperature $T_{\rm dr}$ \cite{cafiero00}.  
The stochastic driving leads thus to an average rate of change 
of temperature 
\begin{equation}
\Delta T_{\rm rot}/\Delta t = H_{\rm dr} ~, 
{\rm ~~ with  ~~} H_{\rm dr} = f_{\rm dr} T_{\rm dr} ~.
\label{eq:Hdr}
\end{equation}

The starting point for our mean-field analysis is the theory of 
Huthmann and Zippelius \cite{huthmann98}, for a freely cooling gas of 
infinitely rough particles,
which was recently complemented by numerical simulations in 2D and
3D \cite{luding98d} and by studies of driven systems as well \cite{cafiero00}. 
The main outcome of this approach is a set of coupled evolution 
equations for the translational and rotational MF temperatures 
$T_{\rm tr}$ and $T_{\rm rot}$ \cite{huthmann98} which can be extended
to describe arbitrary energy input (driving) \cite{cafiero00}. 
In the present study, given the random driving 
temperature $T_{\rm dr}$ and an energy input
rate $f_{\rm dr}$, as defined above, one just has to add the positive 
rate of change of rotational energy $H_{\rm dr}$ to the system 
of equations:
\begin{eqnarray}
\label{mfrp1}
  \frac{d}{dt}  T_{\rm tr}(t) & = & \left[
  - G A T_{\rm tr}^{3/2} + G B  T_{\rm tr}^{1/2} T_{\rm rot}\right] \\
  \frac{d}{dt}  T_{\rm rot}(t) & = & {2} \left[
  G B T_{\rm tr}^{3/2} - G C  T_ {\rm tr}^{1/2} T_{\rm rot}\right] 
                                 + H_{\rm dr}  ~,
\label{mfrp2}
\end{eqnarray}
\begin{figure}
\begin{center}
\epsfig{file=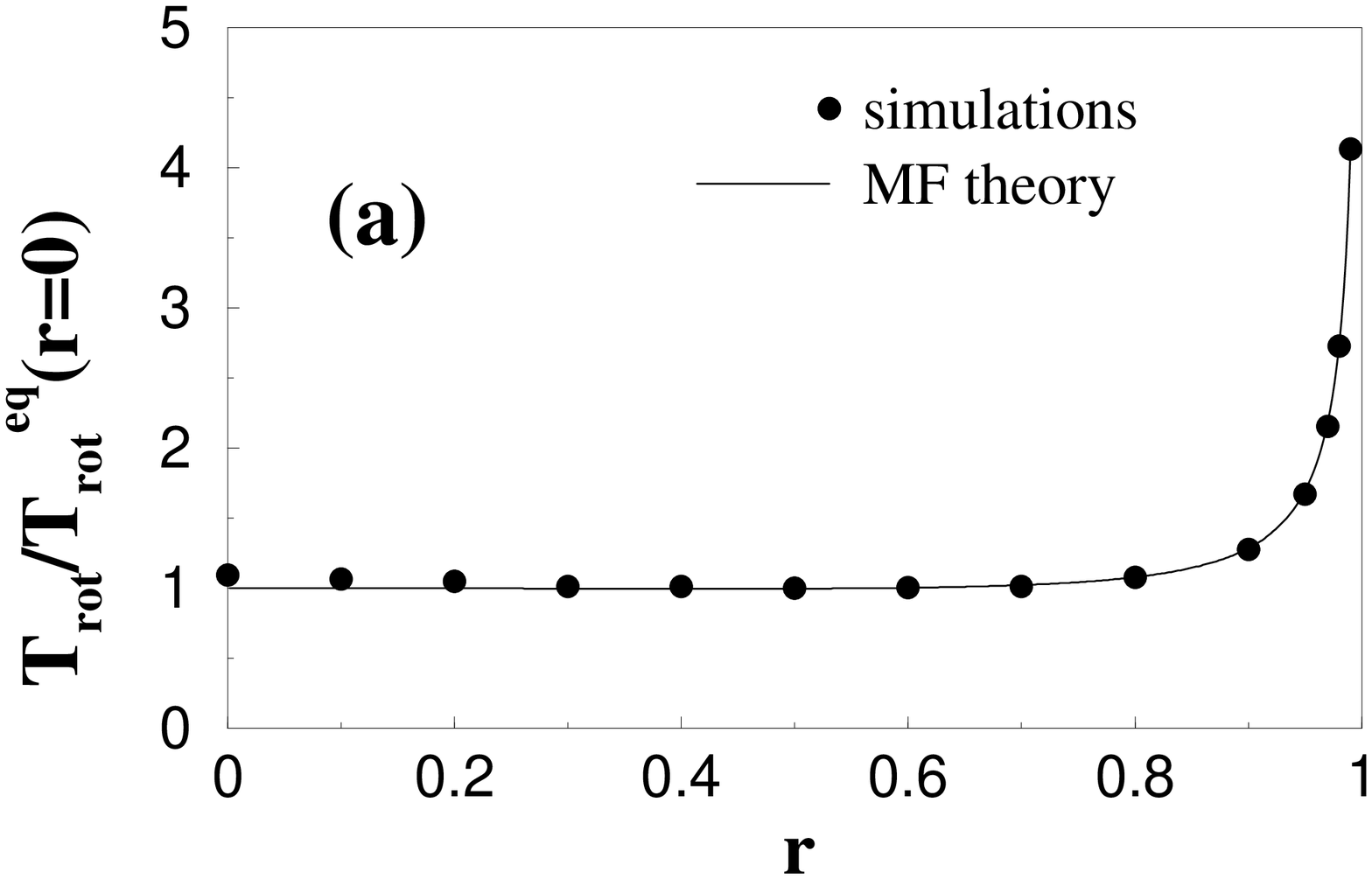,height=4.0cm,angle=0} \\
\vskip -0.2cm
\epsfig{file=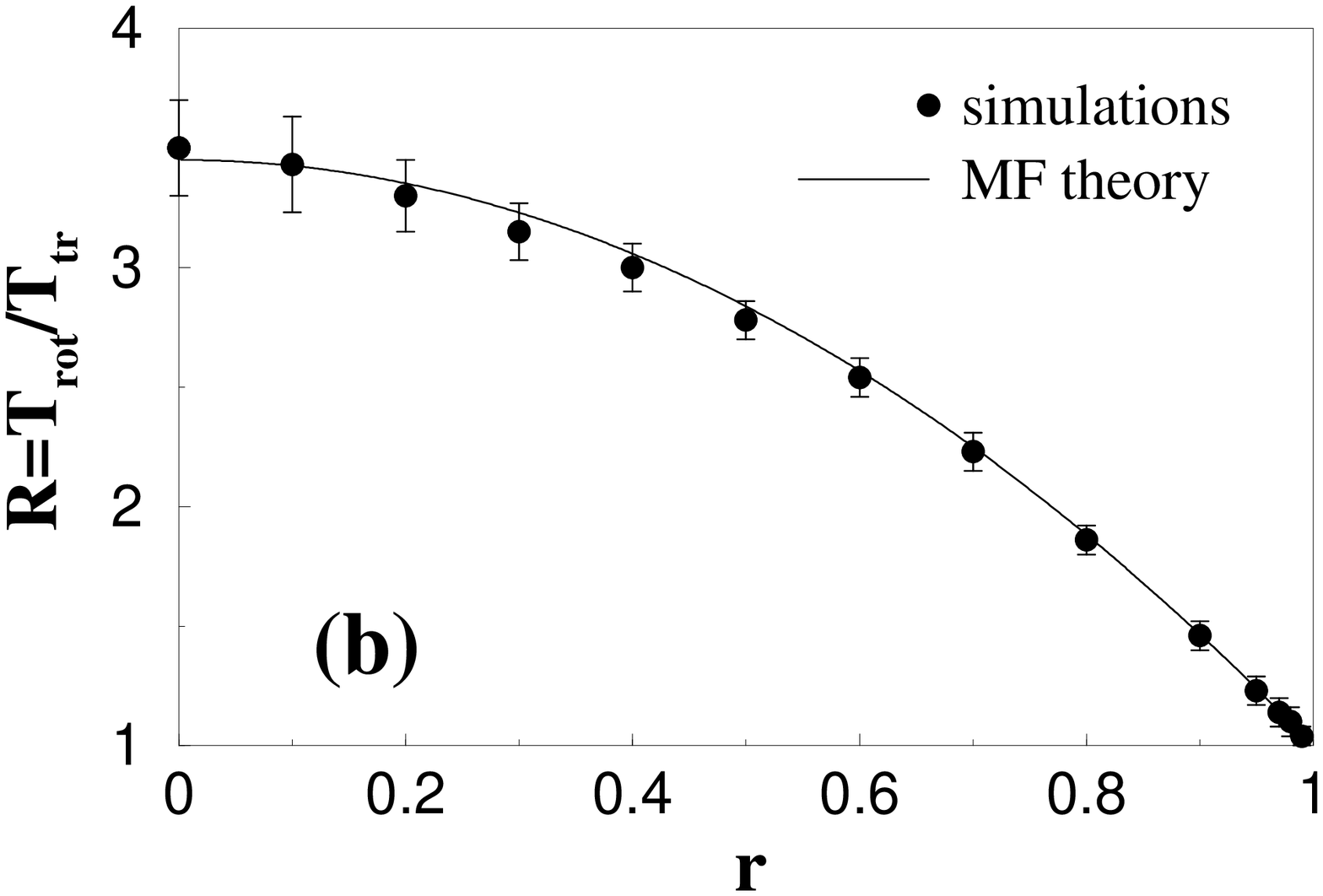,height=4.0cm,angle=0} \\
\vskip -0.4cm
\end{center}
\caption{Simulation (points) and theory (lines) for
the parameters $\nu=0.34$, $N=11025$, and $r_t=1$,
plotted against $r$.
(a) Equilibrium rotational temperature $T_{\rm rot}$, 
normalized by the MF value $T_{rot}^{eq}(r=0)$ at $r=0$.
(b) Ratio of equilibrium rotational and translational temperature
$R=T_{\rm rot}/T_{\rm tr}$.
}
\label{fig2}
\end{figure}

with $G = 8/(\sqrt{\pi M}a) \nu  g_{2a}(\nu)$, and the
pair correlation function at contact $g_{2a}(\nu)=(1-7\nu/16)/(1-\nu)^2$
in the approximation proposed by Henderson
\cite{henderson75,verlet82}, dependent only on the volume 
fraction of the granular gas $\nu=\pi a^2 N/V$.
The constant coefficients in Eqs.\ (\ref{mfrp1}) 
and (\ref{mfrp2}) are $A = {(1-r^2)}/{4} + {\eta}(1-\eta)/2$,
$B = {\eta^2}/(2q)$, and 
$C = {\eta} \left (1-{\eta}/{q}\right ) /(2q)$,
with the abbreviation $\eta=\eta(r_t)={q (1+r_t)}/(2q+2)$,
as derived in Ref.\ \cite{huthmann98},

Setting to zero the temporal derivatives in Eqs.\ (\ref{mfrp1}) 
and (\ref{mfrp2}), one obtains the steady state properties of the 
system:
\begin{equation}
T^{eq}_{\rm rot}=\left( \frac{H_{\rm dr}}{G \Gamma} \right )^{2/3} 
  \,,{\rm ~~and~~} \,
T^{eq}_{\rm tr}  =  T^{eq}_{\rm rot}/R ~,
\label{mfeq}
\end{equation}
with $\Gamma=2\left({B}/{A}\right)^{1/2} 
\left(C-{B^2}/{A}\right)$, and $R=A/B$.

In Fig.\ \ref{fig2} we present equilibrium values of 
$T_{\rm rot}$, normalized by the MF 
value $T_{\rm rot}^{eq}(r=0)$ at $r=0$, 
and of the ratio $R=T^{\rm eq}_{\rm rot}/T^{\rm eq}_{\rm tr}$, as
obtained from numerical simulations of a system 
of $N=11025$ particles, with volume fraction $\nu=0.34$, 
$r_t=1$, and $r$ ranging from $0.99$ to $10^{-4}$. Surprisingly, 
the agreement with the MF prediction is very good, 
even {\em for the lowest value $r=10^{-4}$} of the normal restitution, 
which corresponds to very strong dissipation, were the deviation from MF theory is of the order of only $10\%$.
\begin{figure}[ht]
\begin{center}
\hspace{-0.3cm}
\epsfig{file=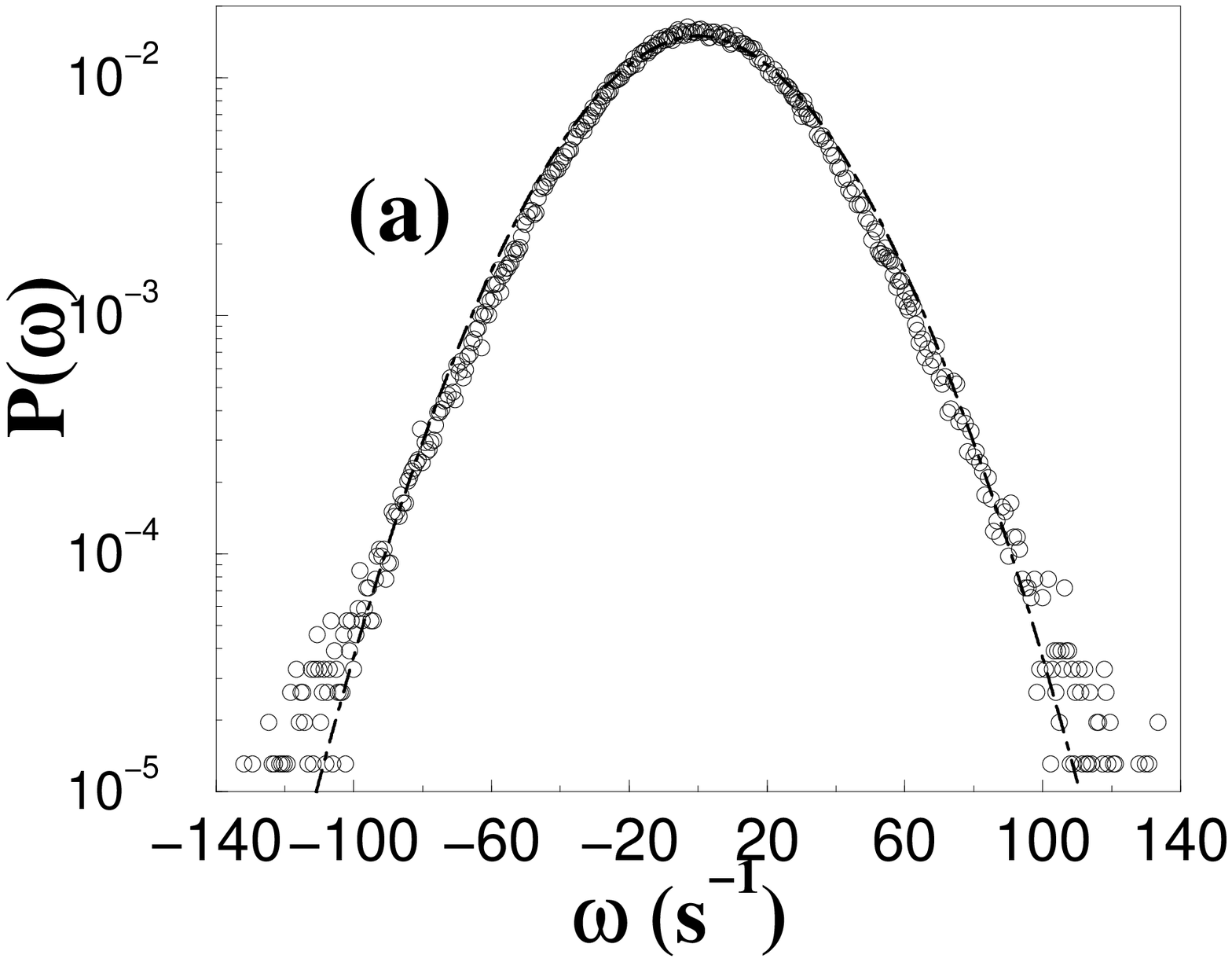,height=4.50cm,angle=0}
\hspace{0.3cm}
\epsfig{file=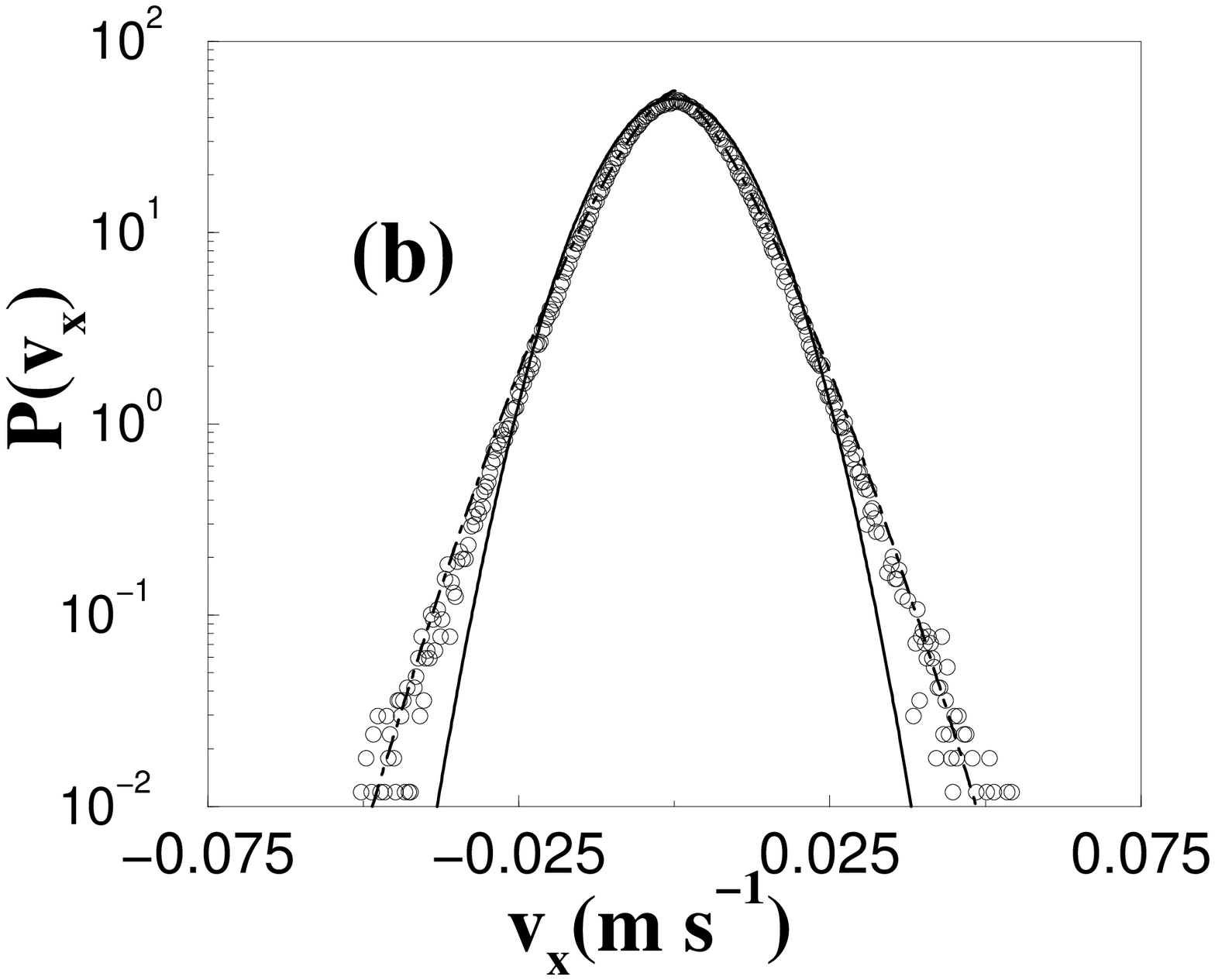,height=4.74cm,angle=0}
\vspace{-0.3cm}
\end{center}
\caption{Steady state rotational (a) and translational (b)
velocity distributions for $N=11025$,
$\nu=0.34$, $r_t=1.0$ and $r=0.1$. A power law fit (dashed line)
gives an exponent $\alpha=1.92(6)$ for the rotational distribution and
$\alpha=1.41(6)$ for the translational distribution (see text for details).
For comparison, a Maxwellian (full line) is plotted in (b).}
\label{fig3}
\end{figure}To give an example, if the system is driven 
on the translational degrees of freedom, the equilibrium 
temperatures show deviations of $30-40\%$ from MF predictions 
already for $r=0.6$, see \cite{cafiero00}. The snapshot in 
Fig.\ \ref{fig2} shows the equilibrium particle distribution for 
$r=0.1$ and appears spatially homogeneous -- apart from small density 
fluctuations not quantified here. 

In Fig.\ \ref{fig3}, we show the equilibrium
rotational and translational velocity distributions for $r=0.1$, 
with the other parameters as above. The rotational velocity 
distribution is very near to a Maxwellian. A three parameter 
fit $f(x)=A \exp(-B | ({x-\langle x\rangle})/{\sigma}|^{\alpha})$, where 
$\sigma=\left(\langle(x-\langle x\rangle)^2\rangle\right)^{1/2}$, 
and $x$ either equals
$\omega$ or $v$, is plotted as dashed line in Fig.\ \ref{fig3}. 
The parameters $\langle\omega\rangle$ and $\sigma$ are taken from 
the simulations, and the fit gives $\alpha=1.92(6)$ for $\omega$,
while we obtain $\alpha=1.41(6)$ for $v$. Rotational velocities 
are then characterized by good homogeneization at low $r$, while
the translational velocity distribution shows strong deviations 
from a Maxwellian.
This deviation is due to the high dissipation. 
Numerical simulations with $r=0.99$ give a Maxwellian distribution 
for both rotational and translational velocities.

\begin{figure}[htb]
\begin{center}
\epsfig{file=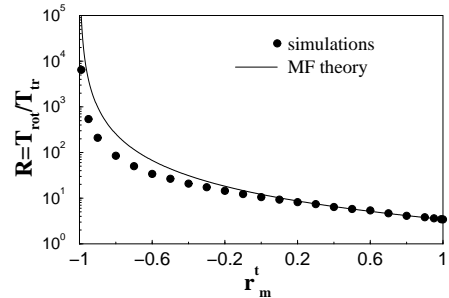,height=4.0cm,angle=0} \\
\vskip -0.4cm
\end{center}
\caption{Simulation (points) and theory (lines) results for
$R=T_{\rm rot}/T_{\rm tr}$, with parameters $\nu=0.34$, $N=11025$,
and $r=0.1$, plotted against $r_t$.}
\label{fig4}
\end{figure}

In order to check the role of the tangential restitution, we show in
Fig.\ \ref{fig4} the equilibrium values of $R$ with $r=0.1$ and 
$r_t \in [-1,1]$. While for positive $r_t$ there is still good agreement 
with MF theory, strong deviations appear as $r_t\to -1$.
Note that many realistic materials obey the relation $r_t \approx 0.4$
\cite{foerster94}, what renders our mean field approach still acceptable.

The conclusions of our study are that the driving on the rotational 
degrees of freedom is able to keep the spatial 
homogeneity of the system up to very high dissipation rates, 
for positive values of $r_t$. 
This leads to a very good agreement of the equilibrium temperatures 
with the MF predictions. There are two 
possible reasons for this. From one side, the driving acts 
on rotations. Then, it cannot favorise collisions, since it does not increase 
the normal component of the relative velocity of the 
colliding particles. From the other side, the increase of 
rotational energy triggered by the driving leads to a shearing force between 
particles, which reduces density fluctuations and destroys 
velocity correlations. When $r_t \to -1$, 
the agreement with MF is lost. To explain this 
result one has to remember that $1+r_t$ is a measure for the
strength of the coupling.  Not enough rotational energy
is transferred to the translational degree, so that the 
randomization on collision does not take place. Thus,
it is not surprising that MF is no more valid in this very singular limit.
Snapshots of the particle distribution for $r=0.1$ and $r_t$ near to $-1$ 
(not displayed here) show indeed stronger density fluctuations in the system
as reported in Fig.\ \ref{fig1}.  

The translational velocity distribution exhibits strong deviations 
from MF prediction, in the {\em homogeneous}
high dissipation regime, showing from one side 
that deviations from a Maxwellian are not necessarily related
to clustering, and from the other side that it is possible 
to have a good agreement of the second moment 
(the temperature) of the velocity distribution with MF theory 
{\em together with} a non Maxwellian velocity distribution. This 
poses a theoretical challenge, since recently proposed theories 
for horizontally driven granular gases \cite{puglisi98}, 
assume that clustering is responsible 
for fat tails in the velocity distribution.

Apparently the {\em temperature} of the system depends
 mainly on the energy balance relations, which depends 
indirectly on density fluctuations (density fluctuations 
influence strongly both the frequency of collision and the rate 
of dissipated energy per collision) while higher moments, and 
the overall shape of the velocity distributions are 
more sensible to other details.

A difficult to realize, but probably good, 
experimental setup is the following. Each, extremely rough,
granular sphere, contains a small (to reduce the effect of
 dipole-dipole interaction at collision) magnetic bar. The plane on which 
the spheres move should be extremely smooth, in order
to avoid energy dissipation. Then, a spatially homogeneous magnetic 
pulses periodically spaced in time can be applied 
in the horizontal directions. 
This would be the magnetic analogon of the oscillating plane. 
If the magnetic field is really spatially homogeneous, 
the magnetic dipoles of the spheres will receive angular 
momentum from the field, so only rotations are driven, 
and this angular momentum will be ``randomized'' by 
the collisions, if they are frequent enough, exactly like it happens 
for the kinetic energy coming from the oscillating 
plane usually employed in experiments. To reach an equilibrium, 
it is necessary to give an initial 
translational velocity to the particles. 
We are aware that such an experiment is extremely difficult to realize, 
but we hope there will be some experimentalist willing to try it.

Summarizing, the main discovery of this work is that a dissipative
gas has strongly anomalous velocity distributions even in the 
absence of large-scale inhomogeneities.  This is achieved by injecting the 
energy into rotational motion and allowing for a transfer to 
translations through strong friction.  The system acts like a
``transformer'' converting Maxwellian degrees of freedom into
distributions with fat tails.

R. C. and  H. J. H. acknowledge financial support 
under the European network project FMRXCT980183; S. L. 
and H. J. H. acknowledge funding from the Deutsche Forschungsgemeinschaft
(DFG). We thank E. Cl\'ement for helpful discussions.


\end{document}